\documentclass[letterpaper, 10 pt, conference]{ieeeconf}  
\IEEEoverridecommandlockouts
\usepackage{cite}[sort,compress]
\usepackage[pdftex]{graphicx}
\usepackage{array}
\usepackage{url}
\usepackage{arydshln}
\usepackage{float}
\usepackage{scrextend}
\usepackage{blindtext}
\usepackage{amsmath,amsfonts,amssymb}
\usepackage{tikz}
\usepackage{xspace}
\usepackage{mathrsfs,mathtools}
\usepackage{mathalfa}
\usepackage{euscript}
\usepackage{csquotes}
\usepackage{datetime}
\usepackage{arydshln}
\usepackage{multirow}

\let\labelindent\relax
\usepackage[inline]{enumitem}

\def\arxivversion{1} %
\newcommand{\arxivver}[2]{%
  \ifx\arxivversion\undefined%
    #1%
  \else%
    #2%
  \fi%
}

\newcommand{\comment}[1]{}
\newcommand{\chris}[1]{{\color{orange}#1}}
\renewcommand{\chris}[1]{#1}
\newcommand{\jaap}[1]{{\color{mgreen}#1}}

\definecolor{mblue}{rgb}{0,0.4470,0.7410}
\definecolor{morange}{rgb}{0.8500,0.3250,0.0980}
\definecolor{myellow}{rgb}{0.9290,0.6940,0.1250}
\definecolor{mpurple}{rgb}{0.4940,0.1840,0.5560}
\definecolor{mgreen}{rgb}{0.4660,0.6740,0.1880}
\definecolor{mcyan}{rgb}{0.3010,0.7450,0.9330}
\definecolor{mred}{rgb}{0.6350,0.0780,0.1840}
\definecolor{mgreenblue}{rgb}{0.0,1.0,0.5}
\definecolor{parulablue}{rgb}{0.2431,0.1490,0.6588}
\definecolor{parulalblue}{RGB}{39,151,235}
\definecolor{parulagreen}{RGB}{129,204,89}
\definecolor{parulayellow}{RGB}{249,251,21}
\definecolor{cblue}{rgb}{0,0.9,1}
\definecolor{corange}{rgb}{1,0.7,0}
\definecolor{mgray}{rgb}{0.8,0.8,0.8}

\usepackage{amsthm}

\theoremstyle{definition}
\newtheorem{defn}{Definition}
\newtheorem{exmp}{Example}
\theoremstyle{plain}
\newtheorem{theorem}{Theorem}
\newtheorem{lem}{Lemma}
\newtheorem{prop}{Proposition}
\newtheorem{cor}{Corollary}

\newtheorem{assumption}{Assumption}
\newtheorem{rem}{Remark}
\theoremstyle{remark}

\newtheorem{cond}{Condition}

\newcommand{\tss}[1]{\textsuperscript{#1}}

\newcounter{ass}

\newcommand{\mc}[1]{\mathcal{#1}}

\newcommand{\mr}[1]{\mathrm{#1}}
\newcommand{\mb}[1]{\mathbb{#1}}
\newcommand{\ms}[1]{\mathscr{#1}}

\newcommand{\mt}[1]{\mathtt{#1}}

\newcommand{\R}{\mathbb{R}}
\newcommand{\dnx}{n_\mr{x}}

\newcommand{\dnu}{n_\mr{u}}
\newcommand{\dnp}{n_\mr{p}}

\newcommand{\dnv}{n_\mr{v}}

\newcommand{\posdef}{\succ}
\newcommand{\negdef}{\prec}
\newcommand{\possemidef}{\succcurlyeq}
\newcommand{\negsemidef}{\preccurlyeq}

\newcommand{\setdefinition}[2]{\left\{\vphantom{#2}#1\right.\left|\,\vphantom{#1}#2\right\}}

\newcommand{\kron}{\otimes} %

\newcommand{\Nd}{{N_\mr{d}}}

\newlength\figH 
\newlength\figW
\usepackage{pgfplots}

\newcommand{\matN}{\Upsilon}
\newcommand{\matM}{\Lambda}

\begin{document}
\title{%
Decoupling parameter variation from noise:\\Biquadratic Lyapunov forms in data-driven LPV control}
\author{%
Chris Verhoek, %
Jaap Eising, Florian D{\"o}rfler, and Roland T{\'o}th %
\thanks{%
This work has been supported by the European Union within the framework of the National Laboratory for Autonomous Systems (RRF-2.3.1-21-2022-00002) as well as the SNF/FW Weave Project 200021E\_20397.
}
\thanks{C.~Verhoek and R.~T{\'o}th are with the Control Systems Group, Eindhoven University of Technology, The Netherlands. J.~Eising and F.~D{\"o}rfler are with the Institut f{\"u}r Automatik, ETH Z{\"u}rich, Switzerland. R.~T{\'o}th is also with the Systems and Control Laboratory, HUN-REN Institute for Computer Science and Control, Hungary. Email addresses: \texttt{\{c.verhoek, r.toth\}@tue.nl}, and \texttt{\{jeising, dorfler\}@ethz.ch}.}
\thanks{Corresponding author: C.~Verhoek.}
}

\maketitle

\begin{abstract}
	A promising step from linear towards nonlinear data-driven control is via the design of controllers for \emph{linear parameter-varying} (LPV) systems, which are linear systems whose parameters are varying along a measurable \emph{scheduling signal}. However, the interplay between uncertainty arising from corrupted data and the parameter-varying nature of these systems impacts the stability analysis and limits the generalization of well-understood data-driven methods available for linear time-invariant systems. In this work, we decouple this interplay using a recently developed variant of the Fundamental Lemma for LPV systems and the concept of data-informativity, in combination with biquadratic Lyapunov forms. Together, these allow us to develop novel linear matrix inequality conditions for the existence of \emph{scheduling-dependent} Lyapunov functions, incorporating the intrinsic nonlinearity. Appealingly, these results are stated purely in terms of the collected data and bounds on the noise, and they are computationally favorable to check.
\vspace{-3pt}
\end{abstract}

\vspace{-2pt}

\section{Introduction} \label{s:intro}\vspace{-2pt}
Linear and model-based controller design, i.e., controller design based on a \emph{known} linear model of the system, is widespread in both theoretical and practical applications. However, engineering systems are becoming more complex with demanding requirements in terms of performance, interconnectivity and energy efficiency. Hence, handling unknown and nonlinear behaviors is utterly important. For such systems having access to `a known model of the system' can be costly and requires expertise.  The wish for (automated) model-free design with theoretically certified guarantees sparked interest in the field of \emph{direct data-driven control}, where the controller is designed \emph{directly} based on measured data. Lately, a paradigm in which the measurements themselves are viewed as a representation of the system behavior has been gaining a lot of attention (see, e.g.,~\cite{CSS}). This avoids the need for an identification step and thus a model\arxivver{.}{~(see~\cite{dorfler2023data} for a discussion on when to use models).} Important concepts within this paradigm are, e.g., Willems' Fundamental Lemma~\cite{WillemsRapisardaMarkovskyMoor2005}, data-informativity~\cite{vanWaardeEising2020_informativity, eising2021informativity} and identifiability~\cite{markovsky2022identifiability}. However, these have been mainly focusing on \emph{linear time-invariant} (LTI) systems, 
apart from some extensions for specific classes~\cite{depersis2023data, alsalti2021data, guo2021data, markovsky2022data} \chris{or approximations~\cite{eising2023cautious, lazar2023basis}} of nonlinear systems. %
In this paper, we show that these ideas can also be applied to \emph{linear parameter-varying} (LPV) systems, which are often used as surrogates of nonlinear systems. See, e.g.,~\cite{formentin2013direct, bao2022overview} for alternative indirect or learning-based data-driven LPV control~methods.

The LPV framework considers systems where a linear input-(state)-output relationship is varying along a so-called \emph{scheduling signal}. This signal is used to capture time-varying and/or nonlinear aspects of the system behavior. Assuming that the scheduling signal is measurable and varying within a bounded set, 
the LPV framework has proven capable of systematically handling a range of complex nonlinear control problems, whilst retaining many of the desirable properties of LTI control, see~\cite{hoffmann2014survey, Toth2010_book} and references therein. The fact that this approach can be used in nonlinear data-driven control has been (experimentally) shown in~\cite{Verhoek2023_DDLPVstatefb_experiment, verhoek2023genNL} using the LPV extension of Willems' Fundamental Lemma~\cite{verhoek2021fundamental}. These works, however, do not consider noise-corrupted data. We aim to overcome this by employing ideas from the \emph{data-informativity} framework. In simple terms, this means that we characterize the set of all LPV systems that could have generated the observed data-set with bounded disturbances. This set is then used to find a Lyapunov function and a controller that guarantee (robust) stability %
against any uncertainty introduced by both the noise and the scheduling variation. The combination of data-informativity and the LPV framework has been considered before~\cite{miller2022data}. However, the results of~\cite{miller2022data} guarantee stability with a \emph{single}, robust (i.e., scheduling-independent) Lyapunov function. 
This means that this single Lyapunov function is required to decrease, i.e., prove stability of the closed-loop LPV system, for all possible variations of the scheduling signal and assumed noise realizations. This can introduce significant conservatism in stability analysis in data-driven controller design and possibly even make these tasks infeasible.
In this work, we decouple the interplay between the uncertainty introduced due to noise in the data and parameter variation along the scheduling signal within the LPV data-informativity framework. We do so by considering Lyapunov functions that are \emph{biquadratic forms}, that is, quadratic in both the state and scheduling signals.

This work contains the following three contributions:
\begin{enumerate}[label={C\arabic*.}, ref={C\arabic*}]
	\item We derive conditions for stability analysis using biquadratic (thus scheduling-dependent) Lyapunov functions for discrete-time LPV systems. In particular, these generalize previously known results. \label{c1}
    \item We use the conditions of~\ref{c1} to formulate LPV control design methods within the data-informativity framework for LPV systems. This yields \emph{linear matrix inequality} (LMI) based conditions that can be efficiently solved in a \emph{semi-definite program} (SDP).
	\item Finally, we demonstrate and compare the increased stability and robustness range resulting from our methods using a nonlinear simulation example.
\end{enumerate} 
In the remainder, Section~\ref{s:preliminaries} describes the system class and the formal problem statement. Section~\ref{s:QMIs} and~\ref{s:analysis} discuss \emph{quadratic matrix inequalities} (QMIs) and  stability analysis with biquadratic Lyapunov functions, respectively, which we use in the formulation of our data-driven synthesis results in Section~\ref{s:mainresults}. We demonstrate the advantages of our method in Section~\ref{s:example} and give the conclusions in Section~\ref{s:conclusion}.

\subsubsection*{Notation}%
The identity matrix is denoted by $I_n\in\mb{R}^{n\times n}$, while $1_{n}$ denotes the vector $\begin{bsmallmatrix} 1 & \cdots & 1 \end{bsmallmatrix}^\top\in\mb{R}^n$. 
The set of real symmetric matrices of size $n\times n$ is denoted by $\mb{S}^n$. 
$X \succ 0$ and $X\prec 0$ ($X \succeq0 $ and $X \preceq 0$) stands for positive/negative (semi) definiteness of a symmetric matrix $X\in\mb{S}^n$, respectively. The Kronecker product of $A\in\mathbb{R}^{n\times m}$ and $B\in\mathbb{R}^{p\times q}$ is denoted by $A\otimes B \in\mathbb{R}^{pm \times qn}$. Block diagonal concatenation of matrices is given by $\mr{blkdiag}$. For $X = \begin{bsmallmatrix}A & B \\ B^\top & C \end{bsmallmatrix}$,  $X\mid A$ is a shorthand taking the Schur complement of $X$ w.r.t. $A$, i.e., $X\mid A=C-B^\top A^{-1} B$. 

\section{Preliminaries}\label{s:preliminaries}
\subsection{System definition}
Consider a discrete-time LPV system with full state observation \chris{that generates the data, which} can be represented~by
\begin{equation}\label{eq:LPVSS}
    x_{k+1}  = A(p_k)x_k + Bu_k + w_k,
\end{equation}
where $k\in\mathbb{Z}$ is the discrete time, $x_k\in\R^{n_\mr{x}}$, $u_k\in\R^{n_\mr{u}}$ and $p_k\in\mathbb{P}\subset\R^{n_\mr{p}}$ are the measurable state, input, and scheduling signals, respectively, and $\mathbb{P}$ is a compact, convex set that defines the range of the scheduling signal. By the nature of the LPV framework, the signal $p_k$ is considered to be accessible to us, and can be either exogenous or endogenous, i.e., independent from $x, u, w$, or composed as, e.g., $p_k:=\psi(x_k)$ for some function $\psi$. The LPV system is disturbed by a $w_k\in\R^{\dnx}$, which can act as a process disturbance or a (colored) noise process. Note that $B$ here is taken constant for the sake of technical convenience. If $B$ is $p_k$-dependent, then it can be made constant %
by state-augmentation or propagating $u$ through an LTI filter. 
The matrix function $A:\mathbb{P}\rightarrow \mathbb{R}^{n_\mathrm{x}\times n_\mathrm{x}}$ is considered to have affine dependency on $p_k$, which is a common assumption in practice, cf.~\cite{hoffmann2014survey, toth2011state},
\begin{equation}\label{eq:LPVdependency}
    A(p_k)=A_0+{\textstyle\sum_{i=1}^{\dnp}}p_{i,k}A_i,
\end{equation}
where $A_i\in\R^{\dnx\times\dnx}$. The affine scheduling dependence of $A(p_k)$ allows us to write the data-generating system~\eqref{eq:LPVSS} as
\begin{equation}\label{eq:LPVaffine}
    x_{k+1} = \mc{A} L_{p_k} x_k + Bu_k +w_k,
\end{equation}
where we define the shorthand 
\begin{equation}\label{def:A,L} \mc{A}:= \begin{bmatrix} A_0 & A_1 & \cdots & A_{\dnp} \end{bmatrix},\quad  L_{p_k} := \begin{bsmallmatrix} 1 \\ p_k \end{bsmallmatrix}\kron I_{\dnx}. \end{equation}

Apart from the structure~\eqref{eq:LPVdependency}, we assume that~\eqref{eq:LPVSS} is unknown. Instead, we have access to $\Nd+1$ measurements of $x, u$, and $p$, where $x$ is a sample path realization with respect to $u,p,$ and the disturbance $w$. Note that this means that for endogenous $p$, the map $\psi$ is known. The measurements are collected in
\begin{subequations}\label{eq:dataset}
\begin{align}
    \Phi & := \begin{bmatrix} L_{p_0} x_0 & \cdots & L_{p_{\Nd-1}} x_{\Nd-1} \\ u_0 & \cdots & u_{\Nd-1} \end{bmatrix}, \\
    X_+&:= \begin{bmatrix} x_1 & \cdots & x_\Nd \end{bmatrix},
\end{align}
\end{subequations}
where $L_{p_k} x_k = \begin{bmatrix} x_k^\top \!&\! (p_k\kron x_k)^\top \end{bmatrix}_{^{^{\text{\normalsize,}}}}^\top$, while
\begin{equation}\label{eq:noise}
    W_- := \begin{bmatrix} w_0 & \cdots & w_{\Nd-1} \end{bmatrix}.
\end{equation}
The amount of information encoded in the data regarding the system dynamics is quantified by the rank of $\Phi$. In particular, if~$\Phi$ has full row-rank, i.e., $\mr{rank}(\Phi) = \dnu + \dnx(1+\dnp)$, it is said that the data is \emph{persistently exciting} (PE)~\cite{verhoek2022direct}.

\subsection{Problem statement}\label{ss:problem}
Given the above, we are interested in data-based LPV controller synthesis for systems of the form~\eqref{eq:LPVaffine} using only measurements of $x,p$ and $u$. More specifically,
we aim at designing a stabilizing LPV state-feedback controller for~\eqref{eq:LPVaffine} using \emph{only} the measured data-set~\eqref{eq:dataset}, a boundedness assumption on the noise signal~\eqref{eq:noise}, and {the range for $p$, i.e., $\mb{P}\subset\mb{R}^{\dnp}$}. In particular, we will take the viewpoint of the data-informativity framework, meaning that a controller is guaranteed to stabilize the true system only if it stabilizes all systems compatible with the data~\eqref{eq:dataset}. %

In order to develop computationally tractable tests, we will make use of the following technical assumptions:
\begin{enumerate}
    \item {The data-generating system~\eqref{eq:LPVSS} admits a realization of the form~\eqref{eq:LPVaffine}.}
    \item {The measured data is PE, i.e., $\Phi$ in~\eqref{eq:dataset} is full row-rank.}
	\item The noise matrix $W_-$ is bounded in terms of a \emph{quadratic matrix inequality} (QMI).
\end{enumerate}
We will discuss the technicalities of these assumptions in the sections that follow.

\section{Quadratic matrix inequalities in control}\label{s:QMIs}
We assume that the disturbances are compatible with a QMI in terms of a finite-time trajectory, compatible with~\eqref{eq:noise}. This means that, for some $\Pi\in\mb{S}^{\dnx+\Nd}$,
\begin{equation}\label{eq:disturbanceQMI}
    \begin{bmatrix} I_{\dnx} \\ W_-^\top \end{bmatrix}^\top \Pi \begin{bmatrix} I_{\dnx} \\ W_-^\top \end{bmatrix}\possemidef 0.
\end{equation}
The feasible sets of such QMIs will play a major role in this work.
Hence, we define, for some $\Psi\in\mb{S}^{q+r}$, the sets
\begin{subequations}\label{eq:setofQMI}
\begin{equation}\label{eq:setofQMI:nonstrict}
    \mc{Z}_\mr{r}(\Psi) := \setdefinition{Z\in\mb{R}^{r\times q}}{\begin{bmatrix} I_q \\ Z \end{bmatrix}^\top \Psi \begin{bmatrix} I_q \\ Z \end{bmatrix}\possemidef0}.
\end{equation}
Similarly, the interior of~\eqref{eq:setofQMI:nonstrict} is defined by the strict variant:
\begin{equation}
    \mc{Z}^+_\mr{r}(\Psi) := \setdefinition{Z\in\mb{R}^{r\times q}}{\begin{bmatrix} I_q \\ Z \end{bmatrix}^\top \Psi \begin{bmatrix} I_q \\ Z \end{bmatrix}\posdef0}.
\end{equation}
\end{subequations}
On the basis of these definitions in~\eqref{eq:disturbanceQMI},~\eqref{eq:setofQMI}, we have that
\begin{equation}\label{eq:Wdef}
    W_-^\top \in \mc{Z}_{\dnx}(\Pi).
\end{equation}
\begin{rem}[On the noise model]\label{rem:requirementsPi}
    For the existence of disturbances $w$, i.e., $\mc{Z}_{\dnx}(\Pi)\neq\varnothing$, we require \mbox{$\Pi\mid\Pi_{22}\possemidef 0$}. Furthermore, to guarantee \emph{bounded} disturbances, we require $\Pi_{22}\negdef 0$. A simple example satisfying both is $\Pi = \mr{blkdiag}(\Omega, -I_{\Nd})$ with $\Omega\possemidef0$. This means that 
    \begin{equation}\label{eq:energyboundW}
        W_-W_-^\top ={\textstyle\sum_{k=0}^{\Nd-1}} w_kw_k^\top \negsemidef \Omega,
    \end{equation}
that is, we have an energy bound on $W_-$ defined by $\Omega$. In~\cite[Sec.~2]{QMIpaper} other properties of such noise models are discussed. In particular, confidence regions of (colored) Gaussian noise fall into this category \cite[Sec.~5.4]{QMIpaper}. 
\end{rem}
\subsection{Set of LPV systems compatible with~\eqref{eq:dataset}}
Note that $\mc{A},B$ parametrize an LPV system of the form~\eqref{eq:LPVaffine} and that $\dnx,\dnu$, and $\dnp$ follow directly from~\eqref{eq:dataset}.
Hence, in line with the data-informativity framework, we then define the set of LPV systems that are compatible with the measurements~\eqref{eq:dataset} under the disturbance model~\eqref{eq:disturbanceQMI}:
\[ X_+ = \begin{bmatrix} \mc{A} & B\end{bmatrix} \Phi +W_- , \quad W_-^\top \in \mc{Z}_{\dnx}(\Pi).\]
By substituting $W_- = X_+-\begin{bmatrix} \mc{A} & B\end{bmatrix} \Phi$ in~\eqref{eq:disturbanceQMI}, we see that all possible $\mc{A},B$ that could have generated the data-set in~\eqref{eq:dataset} are characterized by the QMI 
\begin{equation}\label{eq:sysQMI}
    \begin{bmatrix} I \\ \mc{A}^\top \\ B^\top \end{bmatrix}^\top \underbrace{\begin{bmatrix} I & X_+ \\ 0 & -\Phi \end{bmatrix}%
\Pi \begin{bmatrix} I & X_+ \\ 0 & -\Phi \end{bmatrix}^{\top}}_{=:\matN}
    \begin{bmatrix} I \\ \mc{A}^\top \\ B^\top \end{bmatrix} \possemidef 0,
\end{equation}
where $\matN\in\mb{S}^{\dnu+\dnx(2+\dnp)}$. In shorthand, we can write $\begin{bmatrix} \mc{A} & B\end{bmatrix}^\top \in\mc{Z}_{\dnu+\dnx(1+\dnp)} (\matN)$.
In this paper, we will assume that the true realization of the noise signal satisfies the noise bound. This means that this set is nonempty by default. Clearly, having a conservative noise bound will lead to larger sets of systems, impeding controller design.

\section{Stability analysis of LPV systems}\label{s:analysis}
\subsection{Controller structure}
To stabilize the system~\eqref{eq:LPVSS}, we will design an LPV state-feedback controller. The state and scheduling signals are measurable and thus available for control, and hence, we design a controller
\begin{subequations}\label{eq:controllaw}
	\begin{equation}\label{eq:controllaw:law}
		u_k = K(x_k,p_k).
	\end{equation}
	To preserve the system class, we choose $K:\mb{R}^{\dnx}\times \mb{R}^{\dnp}\to\mb{R}^{\dnu}$ to have affine dependence on $p_k$ and linear dependence on $x_k$:
	\begin{equation}\label{eq:controllaw:dependencyK}
		K(x_k,p_k) = \big( K_0 + {\textstyle\sum_{i=1}^{\dnp}}p_{i,k}K_i\big)x_k = \mc{K}L_{p_k}x_k,
	\end{equation}
	with $\mc{K}:= \begin{bmatrix} K_0 & K_1 & \cdots & K_{\dnp} \end{bmatrix}$.
\end{subequations}
This allows us to write the closed-loop system as
\begin{multline}\label{eq:closed loop} 
	x_{k+1} = (\mc{A}+B\mc{K})\begin{bmatrix} x_k \\ p_k \kron x_k \end{bmatrix} + w_k  \\ =(\mc{A}+B\mc{K})L_{p_k} x_k + w_k.
\end{multline}
{Hence, the closed-loop is of the same system class as~\eqref{eq:LPVSS}. There are a plethora of model-based synthesis methods available for this class of LPV systems. However, in terms of fully data-driven LPV controller synthesis approaches, only~\cite{verhoek2022direct} (considering only noise-free data) and~\cite{miller2022data} (using common Lyapunov functions) are available. We aim to use \emph{biquadratic} Lyapunov functions, making the analysis (and later the synthesis) problem tractable and less conservative.}

\subsection{Biquadratic Lyapunov functions}
For the ease of presentation, let $w_k\equiv0$ for the remainder of this section. In order to guarantee asymptotic stability of LPV systems of the form~\eqref{eq:closed loop} {under arbitrary variation of $p_k\in\mb{P}$}, we aim at finding a \emph{scheduling-dependent} Lyapunov function that ensures that the LPV system is stable. Hence, we are looking for a $V:\mb{R}^{\dnx}\times\mb{P}\mapsto\R_{\geq0}$ that satisfies
\begin{equation}\label{eq:lyapposdef}
    \alpha(\|x\|) \leq V(x,p) \leq \beta(\|x\|),
\end{equation}
where $\alpha,\beta$ are class-$K$ functions, and
\begin{equation}
    V(x_{k+1},p_{k+1})< V(x_k,p_k),
\end{equation}
for all scheduling sequences $p_k, p_{k+1}\in\mb{P}$ and states $x_k,x_{k+1}\in\mb{R}^{\dnx}\setminus \{0\}$ satisfying \eqref{eq:closed loop}. {We want to highlight that if~\eqref{eq:LPVSS} is an \emph{LPV embedding} of a nonlinear system, i.e., the behavior of~\eqref{eq:LPVSS} \emph{embeds} the nonlinear behavior by defining the scheduling as $p_k:=\psi(x_k)$, then asymptotic stability of~\eqref{eq:LPVSS} implies asymptotic stability of the \emph{origin} of the corresponding nonlinear system~\cite{koelewijn2020pitfalls}.}

As the above class of Lyapunov functions is rather general, the tractability of such Lyapunov stability analysis is limited. The usual method of enabling computationally efficient conditions is to assume that the Lyapunov function does not depend on~$p$ and is merely a quadratic form in~$x$, as for LTI systems.
Instead, here we consider a class of \textit{biquadratic} Lyapunov functions. Thus, we choose $V(x,p)$ to have quadratic dependence on both~$x$ and~$p$. W.l.o.g. %
\begin{equation}\label{eq:lyapfunctionclass}
	V(x_k,p_k) = x_k^\top L_{p_k}^\top P L_{p_k} x_k,
\end{equation}
where $P\in\mb{S}^{\dnx(\dnp+1)}$, and {$L_{p_k}$ is as defined in \eqref{def:A,L}.} In order to guarantee \eqref{eq:lyapposdef}, we take $P\posdef0$. 
\begin{rem}[On positive definiteness and SOS] \label{rem:sos}
	Note that~\eqref{eq:lyapposdef} does not only hold if $P\posdef0$. However, it is known that this relaxation is only nonconservative for the case where either $\dnx=2$ or $\dnp =1$ (see, e.g.,~\cite{Choi1975}). As such, this assumption introduces some conservativeness. To alleviate this, one could employ techniques from the \emph{sum-of-squares} (SOS) literature to test whether for a given~$P$ the function $V(x,p)$ is positive definite instead. 
\end{rem}
\begin{rem}[On biquadratic Lyapunov functions]
    The specific quadratic scheduling-dependence introduced for the Lyapunov functions, i.e.,~\eqref{eq:lyapfunctionclass}, has been introduced for continuous-time LPV systems in~\cite{deSouza2006gain}, where they are used to find controllers with better performance. To the best of our knowledge, the extension towards discrete-time LPV systems has not been made so far.
\end{rem} 

Since we will prove asymptotic stability using the existence of such a biquadratic Lyapunov function, we can derive the following sufficient condition for closed-loop stability. 
\begin{lem}\label{lem:relaxation} 
    Given an LPV system~\eqref{eq:closed loop}, and $V(x_k,p_k) = x_k^\top L_{p_k}^\top P L_{p_k} x_k $, where $P\posdef0$. 
	Then,
    \[ V(x_{k+1},p_{k+1})< V(x_k,p_k), \] 
    for all $x_k,x_{k+1}\in\mathbb{R}^{\dnx} \setminus\{0\}$ and $ p_k,p_{k+1}\in\mb{P}$ {satisfying the dynamics \eqref{eq:closed loop},} if for all $p_{k+1}\in\mb{P}$
    \begin{equation}\label{eq:Lyap suff cond}
        P^{-1} - L_{p_{k+1}} (\mc{A}+B\mc{K}) P^{-1} (\mc{A}+B\mc{K})^\top L_{p_{k+1}} ^\top \posdef 0.
    \end{equation}
\end{lem} 
\begin{proof}
	Since $P\posdef0$, we can employ the Schur complement to show that~\eqref{eq:Lyap suff cond} is equivalent to
	\begin{equation}\label{eq:lem1:convex-in-p}
	    \begin{bmatrix} P^{-1} & L_{p_{k+1}} (\mc{A}+B\mc{K}) \\ (\mc{A}+B\mc{K})^\top L_{p_{k+1}} ^\top & P \end{bmatrix} \posdef0.
	\end{equation}
	Repeating the argument with respect to the other block, we have equivalently:
	\begin{equation}\label{eq:standardLyapMI}
	    P - (\mc{A}+B\mc{K})^\top L_{p_{k+1}} ^\top P L_{p_{k+1}} (\mc{A}+B\mc{K}) \posdef0.
	\end{equation} 
	Let $x_k\neq 0$ and $p_k \in\mathbb{P}$, then we have that $x_k^\top L_{p_k}^\top\neq 0$. Therefore, by premultiplying the last inequality with $x_k^\top L_{p_k}^\top$ and postmultiplying with its transpose we obtain: 
	\begin{equation}\label{eq:Lyap in proof} \underbrace{x_k^\top L_{p_k}^\top P L_{p_k}x_k}_{V(x_k,p_k)} -\underbrace{x_{k+1}^\top L_{p_{k+1}} ^\top P L_{p_{k+1}} x_{k+1}}_{V(x_{k+1},p_{k+1})} >0 . \end{equation}
	We can now conclude the lemma.
\end{proof}
Note that condition \eqref{eq:Lyap suff cond} is the \emph{dual} form, which will prove to be useful in the next section. Furthermore,~\eqref{eq:Lyap suff cond} is only sufficient for asymptotic stability. What inhibits this result from being necessary is the last step of the proof, where we derive~\eqref{eq:Lyap in proof} from~\eqref{eq:standardLyapMI}. Analogous to the discussion in Remark~\ref{rem:sos}, the former can hold under weaker conditions than positive definiteness of~\eqref{eq:standardLyapMI}. 
	
{We have just provided tools to analyze asymptotic stability of an LPV system with biquadratic Lyapunov functions under $w_k\equiv0$. In the next section, we will integrate these tools in the data-informativity framework to formulate data-driven LPV controller synthesis conditions that are only dependent on the noisy data-set~\eqref{eq:dataset}.}

\section{Data-driven LPV controller synthesis}\label{s:mainresults}
{Apart from the structural assumptions presented in Section~\ref{ss:problem}, the technical condition in Remark~\ref{rem:requirementsPi} must also~hold:
\begin{assumption}\label{ass:Pi}
 $\Pi$ satisfies $\Pi \mid \Pi_{22} \possemidef 0$ and $\Pi_{22}\negdef0$.
\end{assumption}
Note that this assumption essentially resembles the uncertainty quantification in the design of the data-generation experiment. We are now ready to present our main results.} 
\subsection{Synthesis approach with scheduling-dependent S-Lemma}
To recap our objective, we aim at finding $\mc{K}$ and $P\posdef0$ such that~\eqref{eq:Lyap suff cond} holds for all scheduling variation $p_{k+1}\in\mathbb{P}$ and all systems $\mc{A},B$ for which $\begin{bmatrix} \mc{A} & B\end{bmatrix}^\top \in \mc{Z}_{\dnx(1+\dnp)+\dnu}(\matN)$, i.e., satisfying~\eqref{eq:sysQMI}.

In other words, we want that for every $Z$ and $p_{k+1}\in\mathbb{P}$: 
\[ \begin{bmatrix} I_{\dnx}\\ Z \end{bmatrix}^{\!\top} \! \matN \begin{bmatrix} I_{\dnx} \\ Z \end{bmatrix}\possemidef 0 \Rightarrow  \begin{bmatrix} I_{\dnx(1+\dnp)}\\ ZL_{p_{k+1}}^{\!\top} \end{bmatrix}^\top \! \matM \begin{bmatrix} I_{\dnx(1+\dnp)} \\ ZL_{p_{k+1}}^\top \end{bmatrix}\posdef0, \] 
where 
\[ \vspace{-1pt} \matM = \begin{bmatrix} P^{-1} & 0  \\ 0 & -\begin{bmatrix} I_{\dnx(1+\dnp)} \\ \mc{K}\end{bmatrix} P^{-1} \begin{bmatrix} I_{\dnx(1+\dnp)} \\ \mc{K}\end{bmatrix}^\top \end{bmatrix}. \] 
Even more concisely, we need:\vspace{-1pt}
\begin{equation}\label{eq:to-slemma-or-not-to-slemma} 
    (\mc{Z}_{\dnu+\dnx(1+\dnp)} (\matN) ) L_{p_{k+1}}^\top \subseteq \mc{Z}_{\dnu+\dnx(1+\dnp)}^+(\matM).\vspace{-1pt}
\end{equation}
{In simple terms, we require a linear transformation ($L_{p_{k+1}}^\top$) of the feasible region of a nonstrict QMI to be contained in the feasible region of another strict QMI. To reduce this to a simple QMI inclusion problem, we will employ~\cite[Thm.~3.4]{QMIpaper}. Note that from Assumption~\ref{ass:Pi}, it follows that $\matN_{22}\negdef0$, which satisfies the assumptions of the QMI result~\cite[Thm.~3.4]{QMIpaper} and allows us to conclude:}
\begin{prop}\label{prop:ZNLp is ZNp}
    Let Assumption~\ref{ass:Pi} hold. Then, for any $p_k\in\mb{P}$
    \[ (\mc{Z}_{\dnu+\dnx(1+\dnp)} (\matN) ) L_{p_k}^\top = \mc{Z}_{\dnu+\dnx(1+\dnp)}(\matN_{p_k}), \] 
    where $\matN_{p_k}\in\mb{S}^{\dnu+2\dnx(1+\dnp)}$ is defined as
    \begin{align*}
        \matN_{p_k}:= & \begin{bmatrix} L_{p_k} & 0 \\ 0 & I\end{bmatrix} \matN\begin{bmatrix} L_{p_k}^\top  & 0 \\ 0 & I\end{bmatrix} \\ 
            = & \begin{bmatrix} L_{p_k} & L_{p_k}X_+ \\ 0 & -\Phi \end{bmatrix} \Pi \begin{bmatrix} L_{p_k} & L_{p_k}X_+ \\ 0 & -\Phi \end{bmatrix}^\top. 
    \end{align*}
\end{prop}	 
{Using this, \eqref{eq:to-slemma-or-not-to-slemma} is equivalent to an inclusion between two feasible regions of QMI's. We can now use the Matrix S-Lemma~\cite[Cor.~4.13]{QMIpaper} to resolve this to an LMI and conclude the following synthesis result.
\begin{theorem}\label{thm:slemma}
	Let Assumption~\ref{ass:Pi} hold and fix a scheduling variable $p_{k+1}\in\mathbb{P}$. There exist scalars $\alpha \geq 0$ and $\beta>0$, a positive definite matrix $F\in\mb{S}^{\dnx(1+\dnp)}$, %
	and a matrix $G \in \mathbb{R}^{\dnu \times \dnx (1+\dnp)}$ such that
	\begin{equation}\label{eq:MI_thm2} 
		\begin{bmatrix} F-\beta I\! & 0 &0 &0  \\ 0 & 0& 0 &F  \\ 0 &0 &0 &G \\ 0& \!F & G^\top & F \end{bmatrix} - \alpha\begin{bmatrix} \matN_{p_{k+1}} \!\!& 0 \\ 0 & 0 \end{bmatrix} \!\possemidef 0,
	\end{equation}
	if and only if inequality~\eqref{eq:Lyap suff cond} holds with $\mc{K}=GF^{-1}$ and $P= F^{-1}$ for all systems $\begin{bmatrix} \mc{A} & B\end{bmatrix}^\top \in\mc{Z}_{\dnu+\dnx(1+\dnp)} (\matN)$, that is, for all systems consistent with the data. 
\end{theorem}
\begin{proof} 
    Starting from \eqref{eq:to-slemma-or-not-to-slemma}, we can apply Proposition~\ref{prop:ZNLp is ZNp} followed by the Matrix S-Lemma~\cite[Cor.~4.13]{QMIpaper}. The assumptions of the latter are directly implied by Assumption~\ref{ass:Pi}. This yields that for a given $p_{k+1}\in\mathbb{P}$, we have \eqref{eq:to-slemma-or-not-to-slemma} if and only if, there exists $\alpha\geq 0$ and $\beta>0$ such that
    \[ \matM - \alpha \matN_{p_{k+1}} \possemidef  \begin{bmatrix} \beta I & 0 \\0 &0 \end{bmatrix} . \]
    {By a Schur complement argument, this %
    is equivalent to
   	\[ \begin{bmatrix} P^{-1}-\beta I\! & 0 &0 &0  \\ 0 & 0& 0 &P^{-1}  \\ 0 &0 &0 &\mc{K} P^{-1} \\ 0& \!P^{-1} & P^{-1}\mc{K}^\top & P^{-1} \end{bmatrix} - \alpha \begin{bmatrix} \matN_{p_{k+1}} \!\!& 0 \\ 0 & 0 \end{bmatrix} \!\possemidef 0. 	\] 
    We can now perform the substitutions $\mc{K}=GF^{-1}$ and $P= F^{-1}$ to conclude the theorem. }
\end{proof}
Theorem~\ref{thm:slemma} provides a tractable condition, corresponding to checking the feasibility of an LMI of size $3\dnx(1+\dnp)+\dnu$, to decide whether a given biquadratic Lyapunov function decreases for all systems compatible with the data and a \emph{given} scheduling variable. What remains to fully resolve the data-driven LPV controller synthesis problem of this paper, is to find $P\posdef0$ and $\mc{K}$ such that for \emph{all} scheduling variables there exist $\alpha\geq 0$ and $\beta>0$ for which \eqref{eq:MI_thm2} holds. 
}

\subsection{Computational approaches}\vspace{-2pt}
We now provide a computational approach to solve the synthesis problem for all $p_{k+1}\in\mb{P}$ as a \emph{semi-definite program} (SDP) subject to a finite number of constraints. We require the following assumption to hold.
\begin{assumption}\label{ass:polytope}
    The set $\mb{P}$ is a convex polytope, generated by $\dnv$ vertices, i.e., $\mb{P} = \mr{co}(\{\mt{p}^i\}_{i=1}^{\dnv})$, where $\mr{co}$ denotes the convex hull and $\mt{p}^i\in\mb{R}^{\dnp}$ denotes a vertex of $\mb{P}$.
\end{assumption}
{Note that the matrix inequality of Theorem~\ref{thm:slemma} is linear in the decision variables.} However, the LMI~\eqref{eq:MI_thm2} seems not necessarily \emph{convex} in the scheduling signal $p_{k+1}$. \chris{However, due to the if-and-only-if relationship between~\eqref{eq:MI_thm2} and~\eqref{eq:Lyap suff cond}, we can state the following:
\begin{cor}\label{cor:1}
    Let Assumption~\ref{ass:Pi} and~\ref{ass:polytope} hold. If~\eqref{eq:MI_thm2} holds for the vertices $\mt{p}^i$, $i\in\{1,\dots, n_\mr{v}\}$, then~\eqref{eq:Lyap suff cond} holds with $\mc{K}=GF^{-1}$ and $P= F^{-1}$ for all systems $\begin{bmatrix} \mc{A} & B\end{bmatrix}^\top \in\mc{Z}_{\dnu+\dnx(1+\dnp)} (\matN)$ and all $p_{k+1}\in\mb{P}$.
\end{cor}
\begin{proof}
    Following the proofs of Theorem~\ref{thm:slemma} and Lemma~\ref{lem:relaxation}, we have~\eqref{eq:MI_thm2}~$\Leftrightarrow$~\eqref{eq:lem1:convex-in-p}~$\Leftrightarrow$~\eqref{eq:Lyap suff cond}. Thus, if~\eqref{eq:MI_thm2} holds for all $\mt{p}^i$, then so does~\eqref{eq:lem1:convex-in-p}. In turn, \eqref{eq:lem1:convex-in-p} is convex in $p_{k+1}$, and thus holds for all $p_{k+1}\in \mb{P}$ by the multi-convexity argument. Hence,~\eqref{eq:Lyap suff cond} holds for all $p_{k+1}\in\mb{P}$. 
\end{proof}
}

\begin{rem}[General remarks on Theorem~\ref{thm:slemma} and Corollary~\ref{cor:1}]\label{rem:genremthm}
\hspace*{-0pt}\vspace{-12pt}
\begin{enumerate}[label={\textit{\roman*)}}, ref={Remark~\ref{rem:genremthm}.\roman*}]
    \item \chris{If Assumption~\ref{ass:Pi} and~\ref{ass:polytope} hold, then one can achieve data-driven LPV controller design by: 1)~Gather data from an unknown LPV system, 2)~Construct $\matN_{\mt{p}^i}$ from the data and $\Pi$ for every vertex of $\mb{P}$, 3)~Solve~\eqref{eq:MI_thm2} on the vertices of $\mb{P}$ using any off-the-shelf SDP solver, and 4)~Implement the control law $u_k = GF^{-1}L_{p_k}x_k$.}
    \item If $\mb{P}$ does not satisfy Assumption~\ref{ass:polytope}, but it is convex and bounded, it can be closely approximated by a polytope. For non-convex sets, one can choose to solve~\eqref{eq:MI_thm2} on a dense grid over~$\mb{P}$. This will, however, only guarantee stability in a neighborhood around the grid points.
    \item The LTI result in~\cite{QMIpaper} is recovered when $\dnp=0$.
    \item For $\dnu=0$, we can similarly formulate a data-informativity-based stability analysis problem with biquadratic Lyapunov functions, which follows the exact same lines as the formulation of the synthesis problems.
    \item Comparing the computational complexity of our methods to the methods in~\cite{miller2022data}, we see that our LMI constraints~\eqref{eq:MI_thm2} are significantly larger ($P\in\mb{S}^{\dnx(1+\dnp)}$ compared to $P\in\mb{S}^{\dnx}$ in~\cite{miller2022data}). Hence, the flexibility introduced by the biquadratic Lyapunov form is traded for an increase in computational complexity.\label{rem:comp}
    \arxivver{}{\item \chris{If the matrix $\matN_{p_{k+1}}$ in~\eqref{eq:MI_thm2} is numerically ill conditioned, one can employ the full-block S-procedure~\cite{Sc01} to relax the problem and make it numerically more stable, see Proposition~\ref{thm:comp} in the Appendix. This, however, sacrifices necessity for convexity.}}
    \end{enumerate}
\end{rem}

\section{Example}\label{s:example}
As an example\footnote{Code at: {\tt gitlab.com/releases-c-verhoek/BLF4DDLPV}.\label{fncode}}, we compare our method with a biquadratic Lyapunov function to the proposed method in~\cite{miller2022data} that considers a constant Lyapunov function. We perform the comparison using a nonlinear system $x_{k+1} = f(x_k) + Bu_k + w_k$ (see the code\tss{\ref{fncode}} for the exact definition of $f$) that is embedded as an LPV system of the form~\eqref{eq:LPVSS}. The embedding has the parameters $\dnx=\dnu=\dnp=2$ with 
\begin{align*}
    A_0 & = \begin{bsmallmatrix} 
        \hphantom{-}0.027 & -0.138 \\ \hphantom{-}0.380 & \hphantom{-}0.014  
    \end{bsmallmatrix}, &&& A_1 & = \begin{bsmallmatrix}
        \hphantom{-}0.449 & -0.164 \\ \hphantom{-}0.129 & -0.257 
    \end{bsmallmatrix},\\
    A_2 & = \begin{bsmallmatrix}
        -0.265 & -0.332 \\ -0.090 & -0.059 
    \end{bsmallmatrix}, &&& B & = \begin{bsmallmatrix} 
        \hphantom{-}0.309 & \hphantom{-}0.539 \\ -0.570 & \hphantom{-}0.467 
    \end{bsmallmatrix},
\end{align*}
and the scheduling variable is defined as 
\[ p_k:=\psi(x_k) = \begin{bmatrix} \delta\sin(x_{1,k}) & \delta\cos(x_{2,k}) \end{bmatrix}^\top, \quad \delta>0. \]  
Hence, it follows that $\mb{P}:=[-\delta, \delta]\times[-\delta, \delta]$, which is a compact and convex polytope. The system is disturbed by a noise signal $w_k\sim\ms{U}(-w_{\max}, w_{\max})$.
For this particular example, we take $\delta=5$ and $w_{\max}=0.1$. If we try to compute\footnote{Using the LPVcore MATLAB toolbox~\cite{lpvcore}, see {\tt lpvcore.net}.} the $\ell_2$-gain for the LPV system with LPV state-space representation $(A(p), [B\ w_{\max} I_{\dnx}], I_{\dnx},0)$ with scheduling ranges $\delta=1$ and $\delta=5$, we obtain \chris{$4.4$ and} {\tt NaN}, respectively, i.e., LPVcore concludes that the open-loop LPV system is unstable for $\delta=5$. 

\begin{figure}
    \centering
    \includegraphics[scale=1]{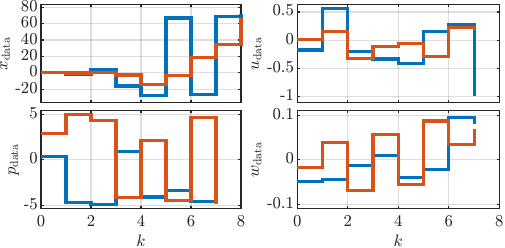}
    
    \vspace{-8pt}
    
    \caption{Data-set used for synthesis. The blue lines represent the first elements of  $x,u,p$ and $w$, while the red lines represent the second elements.}\label{fig:datadictionary}

\vspace{-12pt}

\end{figure}

\begin{figure}
    \centering
    \includegraphics[scale=1]{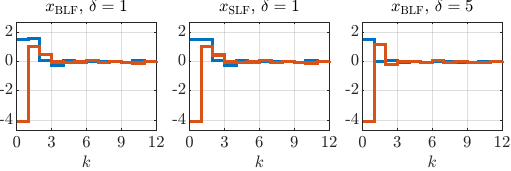}
    
    \vspace{-8pt}
    
    \caption{Closed-loop simulation with the synthesized BLF and SLF based controllers for $\delta=1$ (left two plots) and $\delta=5$ (right plot). The blue lines represent $x_{1,k}$, while the red lines represent $x_{2,k}$. The simulation is performed with a noise signal $w_k\sim\ms{U}(-w_{\max}, w_{\max})$.}\label{fig:sim_jc}
\end{figure}

To obtain our data-set, we simulate the nonlinear system in open-loop with $x_0\sim\ms{N}(0,I_{\dnx})$ for $\Nd=8$ time-steps, where $u_{i,k}\sim\ms{N}(0, 0.5)$, $i=1,2$. The resulting data-set with which we construct~\eqref{eq:dataset} is shown in Fig.~\ref{fig:datadictionary}. With the resulting data-set, $\Phi$ satisfies the  PE condition.
In order to obtain a representation of the disturbance, we choose $\Pi = \mr{blkdiag}(\Omega, -I_{\Nd})$ with $\Omega=10^{-2}\begin{bsmallmatrix} 2.19 & 0.95 \\ 0.95 & 2.63 \end{bsmallmatrix}$, which has been obtained by solving $\min_\Omega \mr{trace}(\Omega)$ subject to~\eqref{eq:energyboundW} for the given $W_-$. Let us abbreviate the synthesis problem of Corollary~\ref{cor:1} by BLF (indicating a \emph{Biquadratic Lyapunov Function}), and the synthesis problem in~\cite[Eq.~(32)]{miller2022data} by SLF (indicating a \emph{Shared Lyapunov Function}). Using only the data-set, the vertices of~$\mb{P}$, and~$\Pi$, we solve the BLF and SLF synthesis problems for both $\delta=1$ and $\delta=5$ using YALMIP with the MOSEK SDP solver in Matlab. Note that SLF returns a polytopic $K(p)$, which can be easily transformed back to the affine form of~\eqref{eq:controllaw:dependencyK} with~\cite[Eq.~(4)]{miller2022data} (see also the code\tss{\ref{fncode}}). For~$\delta=1$, both problems are successfully solved and indeed yield a stable closed-loop response for both BLF and SLF-based controllers, as shown in the left \chris{two plots} in Fig.~\ref{fig:sim_jc}. 
For~$\delta=5$, the BLF synthesis problem is successfully solved \chris{and provides a stabilizing LPV controller, as demonstrated in the simulation in the right plot of Fig.~\ref{fig:sim_jc}.} However, the \chris{SLF synthesis problem could not find a SLF that guaranteed closed-loop stability} for this larger $\mb{P}$. %
This shows that the decoupling of the parameter variation from the noise gives more flexibility (i.e., improved feasibility), at the cost of computational complexity, cf.~\ref{rem:comp}.

\begin{figure}
\centering
    \hfill
    \begin{minipage}[c]{0.4\linewidth}
        \includegraphics[scale=1]{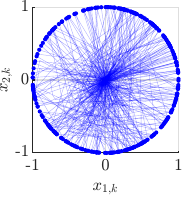}
    \end{minipage}
    \begin{minipage}[c]{0.575\linewidth}\vspace*{45pt}
        \caption{Closed-loop responses with initial conditions on the unit circle of 309 arbitrary systems that are compatible with the data in Fig.~\ref{fig:datadictionary}. Here, we have set $w_k\equiv0$ for the closed-loop simulations.}\label{fig:mult}
    \end{minipage}
    \hfill

    \vspace{-16pt}
\end{figure}

Moreover, note that our analysis proves that the closed-loop corresponding to \emph{any} system compatible with the data in Fig.~\ref{fig:datadictionary} is stable, not just the true system. To illustrate the difficulty, and the different behaviors shown by the systems we stabilize, Fig.~\ref{fig:mult} shows a number of trajectories, 
each corresponding to a random initial condition on the unit circle and a random system compatible with the data in closed-loop with the controller synthesized for $\delta = 5$. This further emphasizes the robustness of the approach.

\section{Conclusions}\label{s:conclusion}
\vspace{-0pt}
In this paper, we developed LPV controller synthesis methods that guarantee Lyapunov stability of the closed-loop system using biquadratic Lyapunov functions, which are less conservative than common Lyapunov functions. The LPV controllers are synthesized using only a single sequence of noisy data, where we assumed that the noise trajectory admits a QMI. A tractable formulation of these results is made possible by the adaptation of the LPV framework in the data-informativity setting, merged with the use of biquadratic Lyapunov functions in LPV controller synthesis. The simulation example demonstrates the increased robustness of the synthesized controllers compared to the controllers synthesized with common Lyapunov functions. In future work, we aim to extend the data-driven state-feedback methods to include performance objectives (e.g., $\ell_2, \mc{H}_2$, passivity, etc.). Furthermore, similar data-driven approaches with output feedback LPV controllers are of interest.

\bibliographystyle{IEEEtran}
\bibliography{references_sfbnoise}

\arxivver{}{\appendix
In this appendix, we provide a methodology to relax the data-driven LPV synthesis problem using the full-block S-procedure~\cite{Sc01}. In particular, we will use the following variant from~\cite{wu2006gain}:
\begin{lem}[Full-block S-procedure~\cite{wu2006gain}]\label{lem:full-block-S-procedure}
	Given the quadratic matrix inequality
	\begin{equation}\label{eq:quad-matrix-inequality}
		L^\top(\mt{p})\Theta L(\mt{p})\negdef0, \quad \forall \mt{p} \in\mathbb{P},
	\end{equation}
	with $L(\mt{p})=\Delta_{\mt{p}}\star \bar{L} =L_{22}+L_{21}\Delta_{\mt{p}}(I-L_{11}\Delta_{\mt{p}})^{-1}L_{12}$, where $\bar{L}= \begin{bsmallmatrix} L_{11} & L_{12} \\ L_{21} & L_{22} \end{bsmallmatrix}$, 
	\begin{equation}\label{eq:deltadef}
		\Delta_{\mt{p}}=\mr{blkdiag}\big(\mt{p}_1 I_{n_{\Delta_1}},\ \mt{p}_2 I_{n_{\Delta_2}},\cdots,\mt{p}_{\dnp} I_{n_{\Delta_{\dnp}}}\big),
	\end{equation}
	and $\mathbb{P}$ is convex. Then, \eqref{eq:quad-matrix-inequality} holds if and only if there exists a real full-block multiplier $\Xi=\Xi^\top$ defined as
	$\Xi=\begin{bsmallmatrix}\Xi_{11} & \Xi_{12} \\ \Xi_{21} & \Xi_{22} \end{bsmallmatrix}$,
	such that
	\begin{subequations}\label{eq:lem:quad-matrix-inequality-multipliers}
		\begin{gather}
			\left[\begin{array}{c c}
				L_{11} & L_{12} \\ I & 0 \\ \hline  L_{21} & L_{22} 
			\end{array}\right]^\top	 
			\left[\begin{array}{c|c}
				\Xi & 0 \\ \hline  0 & \Theta
			\end{array}\right]	
			\left[\begin{array}{c c}
				L_{11} & L_{12} \\ I & 0 \\ \hline  L_{21} & L_{22} 
			\end{array}\right]\prec 0 , \label{eq:lem:quad-matrix-inequality-multipliers:a} \\
			\left[\begin{array}{c}
				I \\ \hline  \Delta_\mt{p}
			\end{array}\right]^\top \Xi
			\left[\begin{array}{c}
				I \\ \hline  \Delta_\mt{p}
			\end{array}\right]  \succcurlyeq 0, \quad\forall \mt{p}\in\mathbb{P}. %
			\label{eq:lem:quad-matrix-inequality-multipliers:b}
		\end{gather}
	\end{subequations}
At the cost of necessity, convexity in $\mt{p}$ can be enforced by adding the additional condition $\Xi_{22}\prec 0$.
\end{lem}
Application of this lemma to Corollary~\ref{cor:1} relaxes the synthesis problem and makes it numerically more stable.
\begin{prop}\label{thm:comp}
    Suppose that Assumptions~\ref{ass:Pi} and~\ref{ass:polytope} hold. Then there exists a positive definite matrix $F\in\mb{S}^{\dnx(1+\dnp)}$, a matrix $G\in\mb{R}^{\dnu\times\dnx(1+\dnp)}$, and scalars $\alpha\geq0$ and $\beta>0$ such that~\eqref{eq:MI_thm2} holds for all $p\in\mathbb{P}$ if there exists $\Xi\in\mb{S}^{2\dnx\dnp}$ with $\Xi_{22}\negdef 0$, and $\epsilon>0$, for which \eqref{eq:lem:quad-matrix-inequality-multipliers} hold for all~$\mt{p}^i$, $i=1,...,\dnv$, where in~\eqref{eq:lem:quad-matrix-inequality-multipliers}:
    \begingroup\allowdisplaybreaks
    \begin{subequations}\label{eq:propcomp:all}
    \begin{align}
        \Theta & = \mr{blkdiag}(\alpha \matN, -(H+\epsilon I)), \label{eq:propcomp:theta}\\ 
        H & = \begin{bmatrix} F-\beta I & 0 &0 &0  \\ 0 & 0& 0 &F  \\ 0 &0 &0 &G \\ 0& F & G^\top & F \end{bmatrix}, \\
        L_{11} & = 0_{\dnx\dnp}, \label{eq:propcomp:L11}\\
        L_{12} & = \begin{bmatrix} 0_{\dnx\dnp\times \dnx} & I_{\dnx\dnp} & 0_{\dnx\dnp\times\dnu+2\dnx(1+\dnp)}\end{bmatrix}, \\
        L_{12} & = \begin{bmatrix} 1_{\dnp}^\top \kron I_{\dnx} \\ 0_{2\dnu+4\dnx(1+\dnp)\times \dnx\dnp} \end{bmatrix}, \\
        L_{22} & = \left[\begin{array}{c} \begin{matrix}\Gamma & 0_{\dnu+\dnx(2+\dnp)\times\dnx(1+\dnp)} \end{matrix} \\\hline I_{\dnu+3\dnx(1+\dnp)} \end{array}\right], \label{eq:propcomp:L22}\\
        \Gamma & = \mr{blkdiag}\left( \begin{bmatrix} I_{\dnx} & 0_{\dnx\times \dnx\dnp} \end{bmatrix}, I_{\dnu+\dnx(1+\dnp)} \right), \\
        \Delta_{\mt{p}^i} &=\mr{blkdiag}\left( \mt{p}^i_1I_{\dnx}, \dots, \mt{p}^i_{\dnp}I_{\dnx} \right).\label{eq:propcomp:delta}
    \end{align}
    \end{subequations}
    \endgroup
\end{prop}
\begin{proof}
    We first write~\eqref{eq:MI_thm2} for some $\mt{p}\in\mb{P}$ in the following quadratic form
    \[ \mc{L}^\top(\mt{p}) \mr{blkdiag}(\alpha \matN,-H)  \mc{L}(\mt{p}) \negsemidef 0, \overset{\exists\epsilon}{\iff} \mc{L}^\top(\mt{p}) \Theta  \mc{L}(\mt{p})\negdef0, \]
    with $\Theta$ as in~\eqref{eq:propcomp:theta} and
    \[ \mc{L}(\mt{p})=\left[\begin{array}{c} \begin{smallmatrix} L_{\mt{p}}^\top & 0 & 0 \\ 0 & I  & 0 \end{smallmatrix}\vspace{.6mm} \\\hline \begin{smallmatrix}I\end{smallmatrix} \end{array}\right]. \]
    We can decompose $\mc{L}(\mt{p})$ as a linear fractional representation~\cite{SchererWeiland2021}, and thus write it as 
    \[ \mc{L}(\mt{p})=L_{22}+L_{21}\Delta_{\mt{p}^i}(I-L_{11}\Delta_{\mt{p}})^{-1}L_{12}, \]
    with $L_{11},\dots, L_{22}, \Delta_{\mt{p}^i}$ as in~\eqref{eq:propcomp:L11}--\eqref{eq:propcomp:delta}. With the quadratic form and the decomposition of~$\mc{L}(\mt{p})$, we can 
    apply the full-block S-procedure of Lemma~\ref{lem:full-block-S-procedure}, which yields \eqref{eq:lem:quad-matrix-inequality-multipliers} with~\eqref{eq:propcomp:all}. By inspection, this condition is linear in the decision variables and convex in~$\mt{p}$, due to $\Xi_{22}\negdef0$. With Assumption~\ref{ass:polytope}, we have a convex condition~\eqref{eq:lem:quad-matrix-inequality-multipliers} over the convex set $\mb{P}$. Hence, solving with~\eqref{eq:propcomp:all} for all $\mt{p}\in\mb{P}$ is equivalent to solving~\eqref{eq:lem:quad-matrix-inequality-multipliers} with~\eqref{eq:propcomp:all} on the vertices $\mt{p}^i$, which concludes the proof.
\end{proof}

}

\end{document}